\documentclass[12pt]{article}

\usepackage{scicite}
\usepackage{graphicx}
\usepackage{txfonts}
\usepackage{hyperref}
\usepackage{breakurl}

\usepackage{times}

\newenvironment{sciabstract}{%
\begin{quote} \bf}
{\end{quote}}

\newcommand{\AN}{\,$\r{A}$}
\newcommand{\kmps}{\,km\,s$^{-1}$}
\newcommand{\rsun}{\,R$_\odot$}
\newcommand{\msun}{\,M$_\odot$}

\newcounter{lastnote}

\title{An unusual white dwarf star may be a surviving remnant of a subluminous Type~Ia supernova}

\author
{S. Vennes$^{1 \ast}$, P. Nemeth$^{2,3}$, A. Kawka$^{1}$, J.R. Thorstensen$^{4}$,\\ 
V. Khalack$^{5}$, L. Ferrario$^{6}$ \& E.H. Alper$^{4}$\\
\\
\normalsize{$^{1}$ Astronomick\'y \'ustav, Akademie v\v{e}d \v{C}esk\'e republiky,}\\ 
\normalsize{Fri\v{c}ova 298, CZ-251 65 Ond\v{r}ejov, Czech Republic.}\\
\normalsize{$^{2}$ Dr. Karl Remeis--Sternwarte, Astronomical Institute,}\\
\normalsize{University Erlangen-N\"{u}rnberg, Sternwartstr. 7, 96049 Bamberg, Germany.}\\
\normalsize{$^{3}$ Astroserver.org, 8533 Malomsok, Hungary}\\
\normalsize{$^{4}$ Department of Physics and Astronomy, 6127 Wilder Laboratory, Dartmouth College,}\\
\normalsize{Hanover, NH 03755-3528, USA.}\\
\normalsize{$^{5}$ D\'epartement de physique et d'astronomie, Universit\'e de Moncton,}\\
\normalsize{Moncton, New Brunswick E1A 3E9, Canada.}\\
\normalsize{$^{6}$ Mathematical Sciences Institute, The Australian National University,}\\
\normalsize{Canberra, ACT 0200, Australia.}\\
\\
\normalsize{${^\ast}$ Corresponding author. E-mail: vennes@asu.cas.cz.}
}

\date{received 23 January 2017; accepted 18 July 2017}

\begin{document} 

\baselineskip24pt

\maketitle 

\begin{sciabstract}

Subluminous Type~Ia supernovae, such as the Type~Iax class prototype SN~2002cx,
are described by a variety of models such as the failed detonation and
partial deflagration of an accreting carbon-oxygen white dwarf star, or the
explosion of an accreting, hybrid carbon-oxygen-neon core.
These models predict that bound remnants survive such events with, according to some
simulations, a high kick velocity. We report the discovery of a high proper motion,
low-mass white dwarf (LP~40-365) that travels at a velocity greater
than the Galactic escape velocity and whose peculiar atmosphere is dominated by
intermediate-mass elements. Strong evidence indicates that this partially burnt remnant was ejected
following a subluminous Type Ia supernova event. This supports the viability of single-degenerate
supernova progenitors.

\end{sciabstract}

Type Ia supernova (SN~Ia) explosions are powered by the detonation of
a Chandrasekhar-mass white dwarf with a degenerate carbon-oxygen core
\cite{nom1997}.
Models show that the explosion may be triggered by
the high internal pressure
caused either by matter accreted from a close donor star [the single degenerate (SD) scenario] or by
the merger with another white dwarf (the double degenerate scenario)
\cite{ibe1984}.
Although Type~Ia supernovae are used to calibrate the cosmological distance scale
\cite{nom1997}
and constrain cosmological models
\cite{rie1998,per1999},
our knowledge of these objects is incomplete and
their progenitors have remained elusive \cite{mao2014,wan2012}.
The possibility of detecting surviving remnants from subluminous SN~Ia events
may help illuminate the SN~Ia progenitor problem in general. 
Models \cite{jor2012,kro2013,fin2014} proposed
to explain observed properties of subluminous SN~Ia such as the SN~Iax-class prototype SN~2002cx
\cite{bra2004,fol2013} involve failed detonation and partial deflagration 
of a massive white dwarf \cite{jor2012,kro2013,fin2014}
or the explosion of a hybrid carbon-oxygen-neon (CONe) core \cite{kro2015,bra2016} 
with both scenarios expected, under the right
circumstances, to leave a bound remnant. Direct evidence for such remnants is missing\cite{mao2014}.

We have observed the high proper motion star LP~40-365\cite{luy1970}.
An identification spectrum was obtained on 2015 February 21
using the Richtey-Chretien spectrograph attached to the Mayall 4-m telescope
at Kitt Peak National Observatory (KPNO) (Fig.~\ref{fig_spec}). The main characteristics are a
blue continuum indicating a temperature of $\approx10\,000$~K (circa B9 star);
the complete absence of neutral hydrogen or helium absorption lines, unlike in normal B stars;
and the appearance of strong magnesium (Mg~{I-II}) and sodium (Na~{I}) line series
and weaker lines of oxygen (O~{I}).
Table~\ref{tbl-par} lists astrometric\cite{lep2005} and photometric\cite{wat2016} data for this object.
We followed up this unusual spectrum using low- to high-dispersion spectra obtained
between June 2015 and June 2016 with the William Herschel 4.2-m telescope on La Palma,
the Hiltner 2.4-m telescope on Kitt Peak, and, finally, with the Gemini-North 8-m
telescope on Mauna Kea\cite{note1}.

We performed a 
spectral line analysis using an iterative procedure that adjusts a parametrized spectral synthesis to the observed line spectrum
using $\chi^2$ minimization techniques. 
These calculations were performed by using a multi-parameter fitting procedure that constrains simultaneously 
the effective temperature ($T_{\rm eff}$) and surface gravity ($\log{g}$) of the star and each individual element abundance in
the atmosphere\cite{note1}. We analyzed the high-resolution 
spectra obtained with the
Echelle SpectroPolarimetric Device
for the Observation of Stars (ESPaDOnS) fed by optical fibers attached to the Gemini-North telescope
using the Gemini Remote Access to CFHT ESPaDOnS Spectrograph (GRACES) (CFHT, Canada-France-Hawaii Telescope). 

\begin{figure}
\hspace{0cm}
\includegraphics[width=1.00\columnwidth]{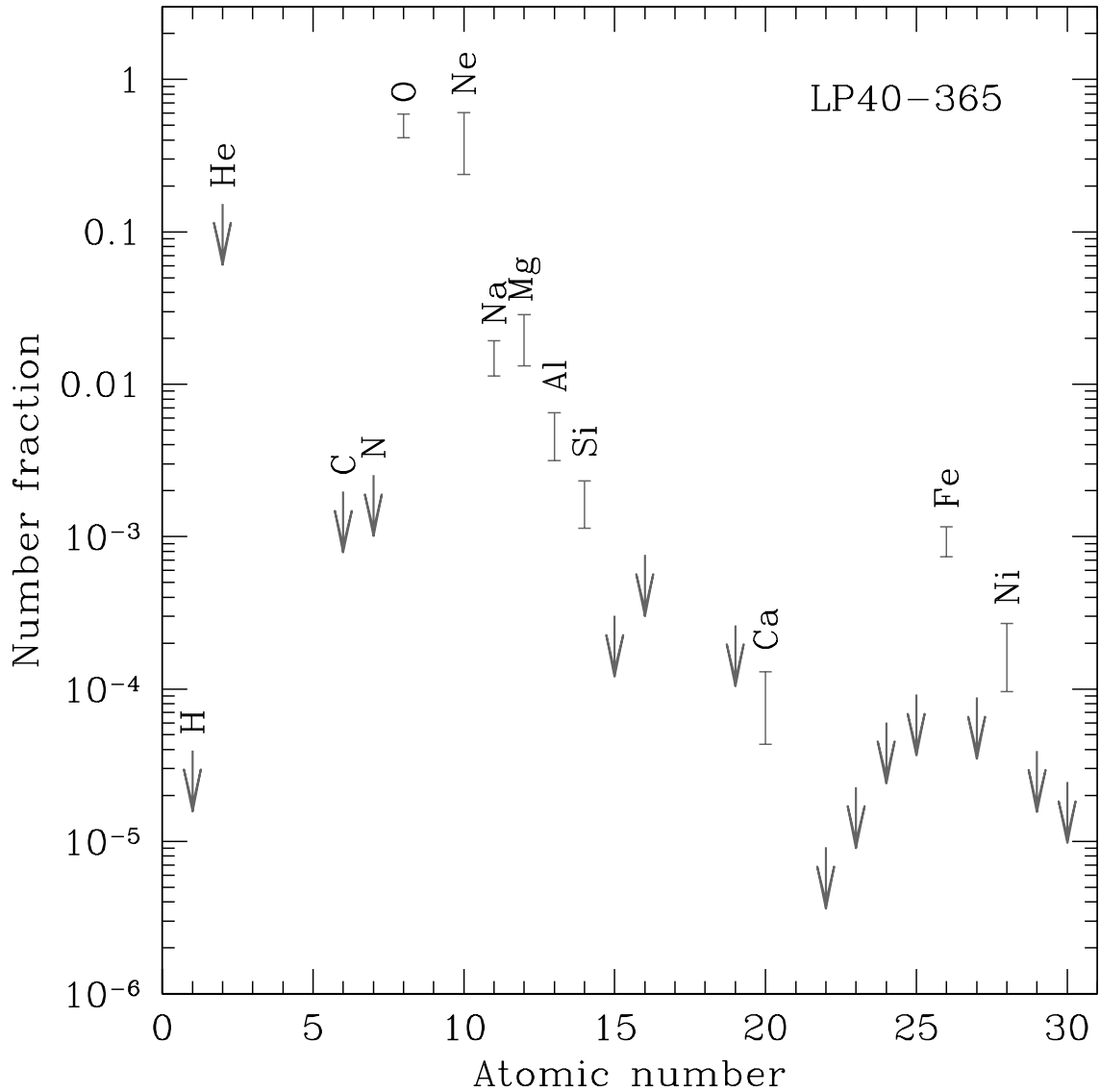}
\caption{{\bf Elemental abundances of LP~40-365.} The photospheric abundances, expressed as the number fraction versus the atomic number, 
were measured in the high-dispersion spectrum
obtained with GRACES
at the Gemini-North telescope on Mauna Kea. The atmosphere is dominated by oxygen and neon followed by sodium and magnesium. 
Iron dominates over nickel and other elements in the iron group by at least a factor of 10. Upper limits 
are shown with arrows.
\label{fig_abun}}
\end{figure}

The model atmospheres and synthetic spectra supporting our analysis were calculated in full non-local thermodynamic equilibrium (non-LTE) 
by using the computer codes {\sc Tlusty} version 204 and {\sc Synspec} version 49\cite{note1,hub1995}.
The chemical composition includes elements with atomic numbers from $Z=1$ (H) to 30 (Zn) with all relevant ionized atoms.
The atmosphere is in radiative equilibrium; convective energy transport was found to be inefficient. 
Detailed line profiles were calculated by using line-broadening parameters dominated by electronic collisions (Stark effect).
Table~\ref{tbl-par} lists best-fitting stellar parameters (with 1$\sigma$ statistical error bars) and Figure~\ref{fig_abun} shows the corresponding abundances.
The high effective temperature and the surface gravity, which is intermediate between normal white dwarfs and the main sequence stars, 
indicate that this object is most likely a low-mass degenerate star\cite{gia2014}.
The line profiles demonstrate that the star is rotating with a projected rotation velocity $v\,\sin{i}=30.5$\kmps, where $i$ is the
apparent inclination of the rotation axis and $v$ is the equatorial rotation velocity, 
suggesting that the parent body was spun up during binary interaction.
The abundance analysis shows that the main atmospheric constituents are oxygen and neon with substantial traces of intermediate-mass 
elements such as aluminium and silicon. 

We measured a large Doppler wavelength shift in the spectral line analysis.
From a sample of 21 velocity measurements taken at different epochs and after correcting for
Earth's motion, we measured an average radial velocity $v_{\rm r}=497.6\pm1.1$\kmps, without significant 
variations ($\chi^2_{\rm r}=1.3$). 
Therefore the star is apparently single and moving at a velocity characteristic of hyper-velocity stars\cite{bro2015}. Those objects
are former members of binary systems that were ejected during three-body encounters with the Galactic center (GC) or 
ejected following the demise of a massive white dwarf
companion in a SD SN~Ia\cite{jus2009}. 

In addition to its large radial velocity, LP~40-365 shows a large apparent motion across the celestial sphere 
of $\mu=158$~milli-arc sec per year (mas\,yr$^{-1}$). With knowledge of the distance
$d$, the proper motion vector
may be converted into the tangential velocity vector $v_{\rm T}$, which combined with the radial velocity $v_{\rm r}$ provides a
complete 3-dimensional description of LP~40-365's motion. 
We estimated the distance toward LP~40-365 with a
photometric method\cite{note1} using as inputs the apparent luminosity of the star and an estimate of its
absolute lumimosity. The absolute luminosity is calculated by using the surface temperature measurements described earlier
and an estimate of the stellar radius which is model dependent. However,
that distance estimate will eventually be superseded by Gaia parallax measurements\cite{per2001}.

The radius of a low-mass, degenerate or partly degenerate body is sensitive to finite-temperature effects\cite{alt1997,ben1999}
that would inflate the radius of a young, extremely low-mass white dwarf such as LP~40-365.
Models for carbon, oxygen, silicon, or iron cores are available\cite{pan2000}, but unfortunately the predicted surface gravity 
of available models ($\log{g}>6.0$, where $g$ is expressed in cm\,s$^{-2}$) largely
exceeds the measured gravity of LP~40-365 indicating that its mass should be much lower than $0.3$ solar mass (\msun).  Lower-mass models with helium
interiors are available\cite{alt1997} and indicate that a body with a mass of $\approx0.14$\msun\ and $\approx8$\% of the solar radius (\rsun) reproduces
the surface gravity and effective temperature of LP~40-365, assuming a cooling age between 5 and 50 million years \cite{note1}. Although we have concluded that the interior of LP~40-365 is most likely composed of
carbon, oxygen and neon, or heavier elements, helium models characterized by identical mean electronic weight ($\mu_e=2$) represent a 
reasonable proxy. The central temperature
of the adopted model, $T_{\rm c}\approx30\times10^6$~K, is lower than that of a typical inert core of normal white dwarfs\cite{alt1997}. Adopting a radius of $0.078$\rsun\ we estimated an absolute magnitude in the Johnson V band $M_{\rm V}=8.14$~mag. Thus, the apparent
$m_{\rm V}$ magnitude listed in Table~\ref{tbl-par} implies a distance of $298^{+150}_{-70}$~pc.
The tangential velocity at a distance of 298~pc is $v_{\rm T}=224$\kmps\ for a
total space velocity relative to the Sun of 546\kmps. 

\begin{figure}
\hspace{0cm}
\includegraphics[width=1.00\columnwidth]{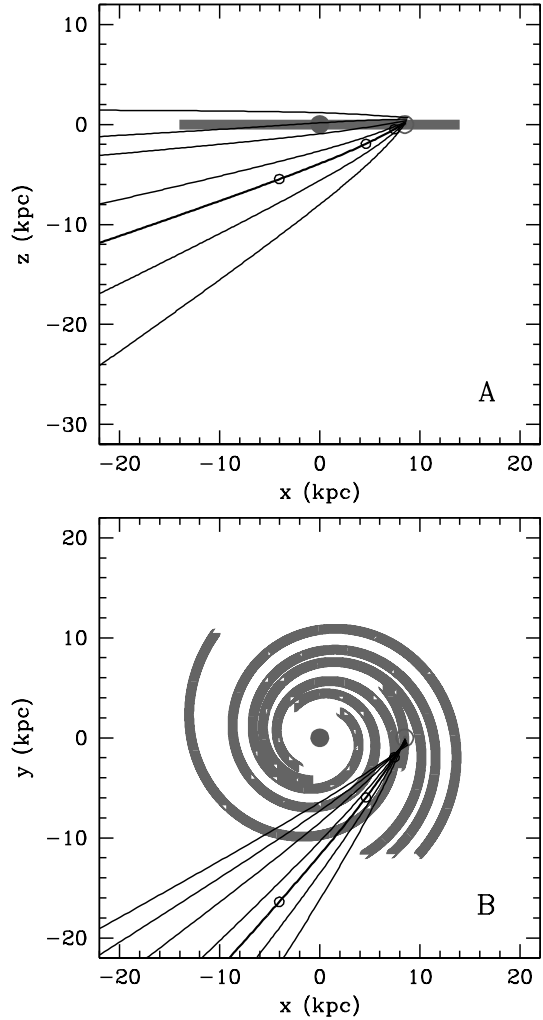}
\caption{{\bf Calculated Galactic motion of LP~40-365.} The orbits are drawn (A) in the Galactic plane ($z$ versus $x$) and (B) perpendicular to the plane ($y$ versus $x$) with the GC (solid circle) at the origin. The Sun ($\odot$) is located 8.5~kpc along the $x$ axis. The current ($t=0$) position of LP~40-365 is estimated assuming a distance to the Sun of, from uppermost to lowermost curve, 1000, 800, 600, 400, 300, 200, and 100~pc. The past trajectory resulting from an assumed distance of 300~pc is marked with open circles at, from rightmost to leftmost circle, $-3$, $-10$ and $-30$ million years. Schematic views of the Galactic arms are shown in gray. 
\label{fig_kine}}
\end{figure}

To retrace the past history of this object, we converted the apparent velocity components (radial and tangential)
into the Galactic velocity vector\cite{joh1987} $(U,\,V,\,W)=(-346,\,360,\,217)$~\kmps. This instantaneous 
velocity vector may be projected back in time 
by adopting an appropriate Galactic potential model\cite{all1991}. We followed the Galactic orbit of LP~40-365 from the present time ($t=0$) back 
to $t=-100$ million years. The projected trajectories displayed in Fig.~\ref{fig_kine} indicate that, for an assumed starting point set at
distances between 100 and 1000~pc, the object did not encounter the GC and, therefore, is not the product of a
three-body dynamical interaction with the GC \cite{hil1988}. None of the resultant trajectories, which allowed for
uncertainties in the distance, are bound Galactic orbits either. The
total velocity in the Galactic rest frame varies between 675\kmps\ assuming $d=100$~pc and 1016\kmps\ assuming $d=1000$~pc with
a velocity of 709\kmps\ at the distance (298~pc) set by the photometric method. All exceed the Galactic escape velocity at 8.5~kpc from the GC\cite{bro2015}.
The object must have originated along one of those projected trajectories and
the trajectory that took LP~40-365 to the present-day distance of 298~pc
entered the plane $< $5 million years ago.
The simulated cooling time scale for a $0.15$\msun\ compact object with an effective temperature of $10^4$~K is only $\approx$5 to 50 million years
\cite{alt1997}. 

Combining the peculiar surface composition of this compact object, the results of the trajectory analysis, 
and the evolutionary age estimate,
it appears likely that LP~40-365 is the surviving remnant of a subluminous SN~Ia event that took place below the Galactic plane, 
a few kiloparsecs away and earlier than 50 million years ago.
The stellar properties and the kinematics of LP~40-365 are comparable to some simulated events\cite{jor2012},
suggesting that this object is indeed a fragment that survived the failed detonation of an SN~Iax.
The mass estimate is somewhat less
than accounted for in these simulations ($> 0.3$\msun). However, other models \cite{fin2014} 
successfully achieve remnants with masses as low
as 0.09\msun\ but without delivering a large kick velocity. Variations in the adopted ignition geometry, such as centered
versus off-center ignition, may affect the kinematical outcome for the surviving remnant.
Simulations involving hybrid 
CONe cores\cite{bra2016} successfully generated low-mass remnants but these simulations did not explore post-explosion kinematics.

Intermediate-mass elements detected in the atmosphere of LP~40-365 are expected to contaminate bound remnants after
a typical SN~Iax event \cite{fin2014}, but
we found only minute traces of iron-group elements which normally dominate the supernova ejecta. 
The paucity of iron-group elements and the prevalence of
lighter elements indicate that gravitational settling and chemical separation may have occurred with light elements dominating
over heavier ones. It is not possible to estimate the fraction of iron material produced in the explosion that would manage to diffuse
to the star's center. Diffusion time scales at the star's surface may be comparable to or longer than the age of the object\cite{paq1986}.
Conversely, the absence of carbon and prevalence of oxygen and neon at the surface of LP~40-365 
would preferably match the configuration of a hybrid CONe core \cite{bra2016}.
None of the simulations take element diffusion explicitly into account, therefore 
a detailed comparison of predicted and observed surface compositions would not constitute a definitive test for any models.

It has been suggested that dynamical instability in a low-mass x-ray binary orbiting a distant main-sequence star could
result in the high-velocity ejection of the donor star
\cite{fre2011,por2011}. 
Apart from an unspecified surface composition,
the predicted high-velocity star could resemble LP~40-365. However, only $\sim10^{-8}$ such events 
are expected per year in the Milky Way \cite{por2011}
compared to a rate of $\sim10^{-3}$ for SN~Iax events \cite{fol2014}; therefore this scenario is less probable.

The actual donor star that must also have been ejected\cite{jus2009} along with LP~40-365 should be detectable as well. 
For example, the high-velocity, helium-rich subdwarf star US~708\cite{gei2015} is a representative
of the class of donor stars that emerged from a SD SN~Ia event and a similar object would have been ejected along
with LP~40-365 after the proposed underluminous SN~Ia event.
The possible detection of a bound remnant in
the aftermath of the SN~Iax event SN~2008ha has been reported although
it may be a chance alignment\cite{fol2014}. The properties of that object are unknown.
The tentative progenitor of SN~2012Z has been described as nova-like \cite{mcc2014} suggesting the likely presence of an accreting white
dwarf in a SD progenitor system akin to that of LP~40-365. No bound remnant has been identified. 
The atmospheric properties of LP~40-365 share some similarities with those of another extreme white dwarf \cite{kep2016} but
exhibit clear distinctions as well: Both WD~1238+674 and LP~40-365 are oxygen-rich but WD~1238+674
is more massive (0.6\msun\ versus 0.14\msun) and its kinematical properties do not appear as extreme. 
The discovery of the oxygen-neon white dwarf WD~1238+674 lends support to the hybrid CONe formation model\cite{doh2015}
and, indirectly, to the subluminous SNIa explosion models involving hybrid CONe white dwarfs\cite{bra2016}.

\clearpage

\bibliography{scibib}

\bibliographystyle{Science}

\vspace{-0.5cm}
\section*{Acknowledgments}

A.K. and S.V. acknowledge support from the Grant Agency of the Czech Republic
(15-15943S). A.K. was a visiting astronomer at KPNO, National Optical Astronomy Observatory.
This work was also supported by the project RVO:67985815 in the Czech Republic.
The research leading to these results has received funding from the
European Community's Seventh Framework Programme (FP7/2013-2016) under
grant agreement number 312430 [Optical Infrared Coordination Network for Astronomy (OPTICON)].
J.R.T. acknowledges support from NSF grant AST-1008217.
This research made use of services at Astroserver.org under the reference number NWKZKA.

This research is based on observations obtained at KPNO, National Optical Astronomy Observatory,
and at the Gemini Observatory, program GN-2016A-FT-24, using ESPaDOnS, located at the
CFHT.
This publication makes use of data products from the Wide-field Infrared Survey
Explorer, 
the Two Micron All Sky Survey, and
the American Association of Variable Sky Observers (AAVSO)
Photometric All Sky Survey (APASS).
Complete acknowledgments are offered as supplementary materials on Science Online.
Observational
data are available in publicly available archives, as are our calculated model
atmospheres. Full details and URLs are given in the supplementary material.

\section*{Supplementary materials}
Materials and Methods\\
Supplementary Text\\
Figs. S1 to S8\\
Tables S1 and S2\\
References (37-58)

\clearpage

\begin{table}
\caption{{\bf Stellar data and parameters.} The celestial coordinates are provided along the right
ascension (RA$\equiv\alpha$) and declination (Dec$\equiv\delta$) and the apparent motion of the star, i.e., the proper motion $\mu$, is decomposed 
into $\mu_\alpha\cos{\delta}$ along the right ascension and $\mu_\delta$ along the declination.\label{tbl-par}}
\centering
\begin{tabular}{lc}
\hline\hline
Parameter & Measurement \\
\hline
RA (J2000)     & 14$^{\rm h}$ 06$^{\rm m}$ 35$^{\rm s}.45$ \\
Dec (J2000)     & +74$^\circ$ 18$'$ 58$''.0$  \\
$\mu_\alpha\cos{\delta}$ & $-56\pm7$ mas\,year$^{-1}$ \\
$\mu_\delta$             & $148\pm7$ mas\,year$^{-1}$ \\
$m_{\rm V}$               & 15.51$\pm$0.09 mag   \\
$T_{\rm eff}$     & $10\,100^{+250}_{-350}$~K  \\
$\log{g/({\rm cm\,s^{-2}})}$         & $5.80^{+0.20}_{-0.35}$ \\
Mass              & $0.14\pm0.01$\msun \\
Radius            & $0.078^{+0.040}_{-0.020}$\rsun \\
$M_{\rm V}$             & $8.14^{+0.60}_{-0.90}$ mag    \\
$v\,\sin{i}$    & $30.5\pm2.0$\kmps\ \\
\hline
\end{tabular}
\end{table}

\begin{figure}
\hspace{0cm}
\includegraphics[width=1.00\columnwidth]{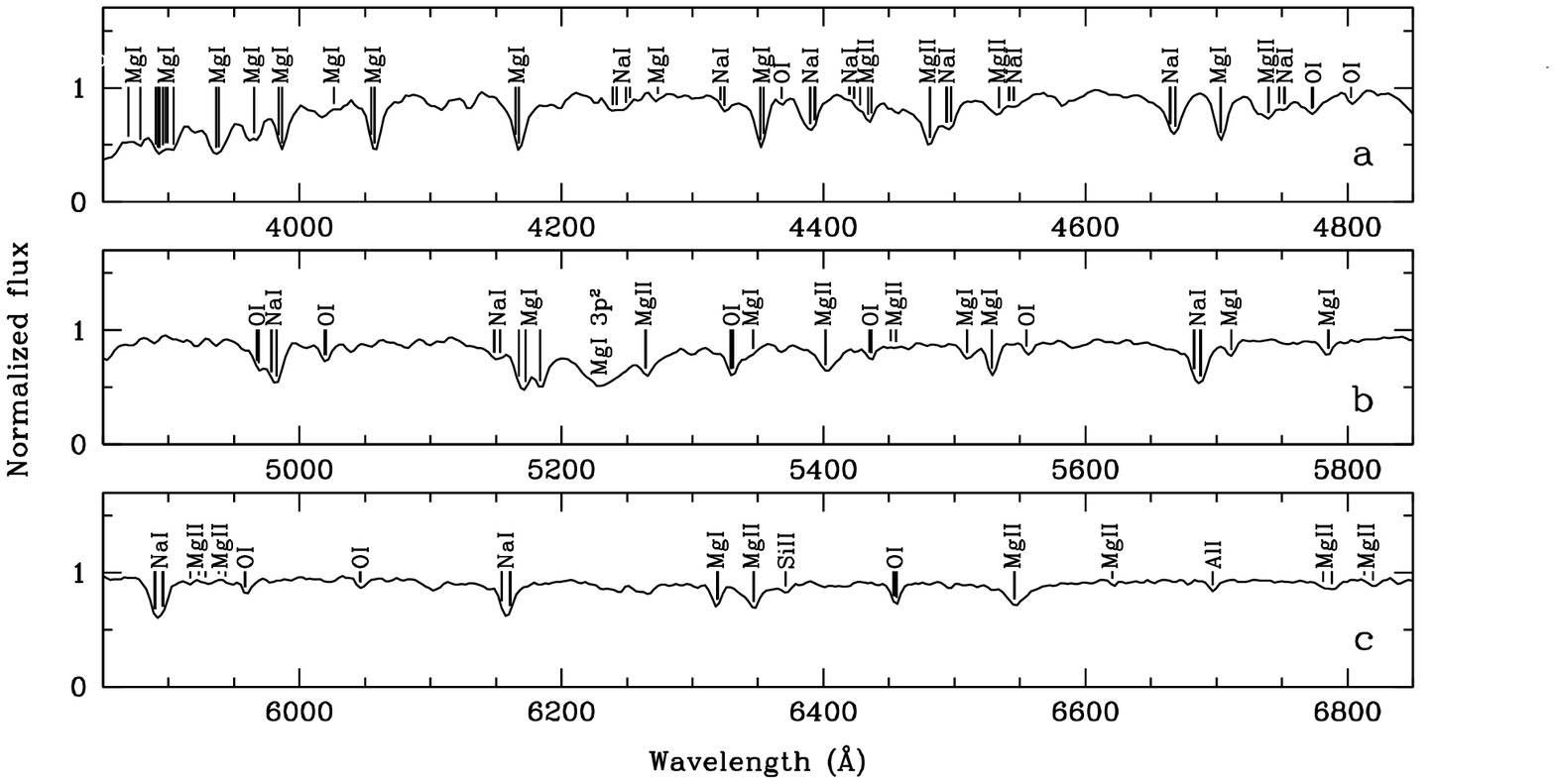}
\caption{{\bf Normalized optical spectrum of LP~40-365 as a function of wavelength.} 
The spectrum was obtained with the Richey-Chretien Spectrograph at the 4-m telescope (KPNO). 
Dominant spectral lines of sodium (Na~{I}) and magnesium (Mg~{I} and Mg~{II}) are labelled along
with weaker lines of oxygen (O~{I}), aluminium (Al~{I}) and silicon (Si~{II}).
A broad feature near 5230\AN\ is tentatively identified with a resonance in the 
Mg~{I}\,3p$^2$ photoionization cross section.
\label{fig_spec}}
\end{figure}

\clearpage

\setcounter{figure}{0}
\setcounter{table}{0}
\setcounter{page}{1}

\renewcommand{\thefigure}{S\arabic{figure}}
\renewcommand{\thetable}{S\arabic{table}}
\renewcommand{\theequation}{S\arabic{equation}}

\begin{center}

{\Large Supplementary Materials for}\\

{\large \bf An unusual white dwarf star may be a surviving remnant of a subluminous Type~Ia supernova}\\

{\large S. Vennes, P. Nemeth, A. Kawka, J.R. Thorstensen, V. Khalack, L. Ferrario, E.H. Alper}\\
\end{center}

\section*{Supporting Online Material}

{\Large\bf  Materials and Methods}\\
\noindent {\bf S1: Spectroscopic observations and radial velocity measurements}\\

In the course of an on-going spectroscopic survey of sub-luminous, 
high proper motion stars\cite{kaw2006,kaw2012},
we uncovered a most peculiar object characterized by stellar parameters 
that would place it below normal
main-sequence late-B stars 
but with a spectrum devoid of hydrogen and helium.

We first observed LP~40-365 on universal time (UT) 2015 February 21 using the Ritchey-Chretien Focus Spectrograph (RC-spec)
\cite{kpno1}
attached to the 4-m telescope at Kitt Peak National Observatory (KPNO).
We employed the T2KA charge-coupled device (CCD) and the KPC-10A grating (316 lines mm$^{-1}$) with a dispersion
of 2.75\AN\ pixel$^{-1}$ in first order and centered on 5300\AN.  We also inserted the
order-sorting filter WG360 and covered a useful spectral range from 3600 to 7200\AN.
The slit width was set at 1.5~arcsecond resulting in a spectral resolution of $\approx5.5$\AN.
We obtained one exposure of 1800~s immediately followed by a second exposure of 1200~s. The observations were conducted
near meridian at an airmass $\approx 1.37$ with the slit oriented East-West (position angle $=90^\circ$) and away from parallactic angle ($=160^\circ$). The resulting differential slit-loss
due to atmospheric dispersion adversely affected the reliability of the flux calibration in the blue part of the
spectrum.

Next, we observed LP~40-365 on UT 2015 June 15-16 with the Intermediate
dispersion Spectrograph and Imaging System (ISIS)\cite{isis1}
mounted on the 4.2-m William
Herschel Telescope (WHT). We employed the standard dichroic (5300\AN) to
separate and direct the light into the blue and red arms.
We used the 1200B and 1200R gratings in the blue and red arms,
respectively, and set the slit width to 1~arcsecond  providing spectral resolutions of $R=5300$ and
$R=9300$, respectively. We obtained five exposures of 1200 s each spread
over two nights to search for variability.

We obtained a set of six spectra from UT 2015 June 23 to July 2 and four additional spectra
on UT 2016 Jan 16, using the 2.4-m Hiltner telescope and
modular spectrograph (Modspec)\cite{mdm1}
at the MDM Observatory on Kitt Peak. We also obtained two additional
spectra on UT 2016 Mar 15 using Modspec attached to the MDM 1.3-m McGraw-Hill telescope. A $2048^2$ SITe CCD gave
2\AN\ pixel$^{-1}$ and, with the slit width set at 1.1~arcsecond at the 2.4-m or 1.8~arcsecond at the 1.3-m telescope,
provided a 3.5\AN\ resolution from 4225 to 7560\AN.

Finally, LP~40-365 was observed with the
Echelle SpectroPolarimetric Device for the Observation of Stars (ESPaDOnS)\cite{man2003} 
in the fast-track service mode offered by the Gemini observatory
on 2016 June 1 and 4 using the Gemini Remote Access to CFHT ESPaDOnS Spectrograph (GRACES)\cite{gemini1}.
We adopted the 2-fiber mode which allows for sky-subtraction with a projected width
on the sky of 1.2 arcsecond for each fiber. This
mode provides a dispersion of 2.88 pixels per resolution element resulting
in a resolving power $R=\lambda/\Delta\lambda\approx40\,000$ in the 2-fiber mode. 
The spectrograph nominally covers a range of 6800\AN\ between 3700 and 10500\AN,
but we obtained useful data with a signal-to-noise ratio exceeding 40
for 21 orders covering 3750\AN\ between 4750 and 8500\AN.
We obtained two exposures of 1955~s each at an airmass of $\approx$1.75.

Table~\ref{tbl_vel} lists individual velocity measurements.
To obtain the measurements in the low- to intermediate-dispersion data we used the {\sc rvidlines} routine in the {\sc rv} package within IRAF\cite{iraf1}.
We employed a set of 15 strong
Mg~{I}, Mg~{II}, and Na~{I} spectral lines
and applied the method to 19 individual spectra obtained with the RC-spec, ISIS/blue and Modspec spectrographs.
All 15 spectral lines from the set were employed to measure the velocities with RC-spec, while only 10 and 7 suitable lines were employed
with ISIS/blue and Modspec, respectively.
Most spectral lines are well separated in the high-dispersion spectra obtained with GRACES. We cross-correlated the spectra with
the best-fitting spectral synthesis using the IRAF routine {\sc fxcor}. 
The weighted average and error of the barycentric-corrected radial velocity measurements
are given by:
\begin{equation}
\bar{v_{\rm r}}=\frac{\sum_{i=1}^{N} w_i\, v_{{\rm r},i}}{\sum_{i=1}^{N} w_i} = 497.6\ {\rm km\,s^{-1}},
\end{equation}
\begin{equation}
\sigma_{v_{\rm r}}=\Big{(}\sum_{i=1}^{N} w_i\Big{)}^{-1/2} = 1.1\ {\rm km\,s^{-1}},
\end{equation}
where $w_i=1/\sigma_{v,i}^2$ and $N=21$.
The reduced $\chi^2$,
\begin{equation}
\chi^2_{\rm r} = \frac{1}{N-1} \sum_{i=1}^{N}w_i\,(v_{{\rm r},i}-\bar{v_{\rm r}})^2=1.3,
\end{equation}
indicates that the radial velocity does not vary significantly.
Including only higher precision measurements (WHT and Gemini) increases the
reduced $\chi_{\rm r}^2$ to $\approx$2.0 which reveals the effect of small systematic errors of $\approx 3$\kmps\
in addition to small statistical errors.\\

\noindent {\bf S2: Spectral atlas}\\

\begin{figure}
\hspace{0cm}
\includegraphics[width=1.00\columnwidth]{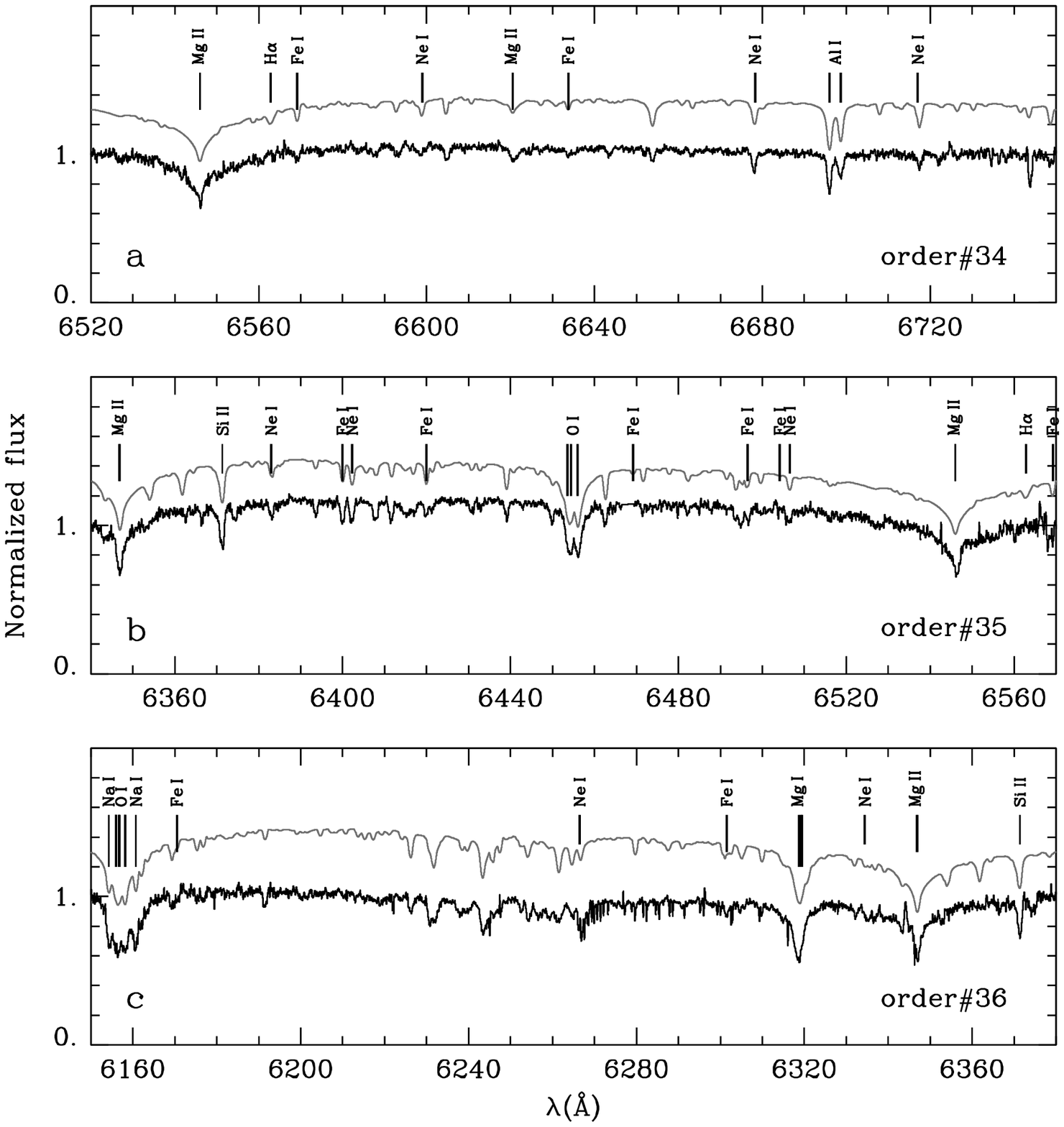}
\caption{{\bf High dispersion spectra and model.} Gemini-North high-dispersion spectra covering orders 34 to 36 (lower lines) and
normalized to unity. The
best-fitting model spectra are shown in grey (top lines) are offset for clarity by $+0.3$. Main spectral features are marked.
\label{fig_atl1}}
\end{figure}

\begin{figure}
\hspace{0cm}
\includegraphics[width=1.00\columnwidth]{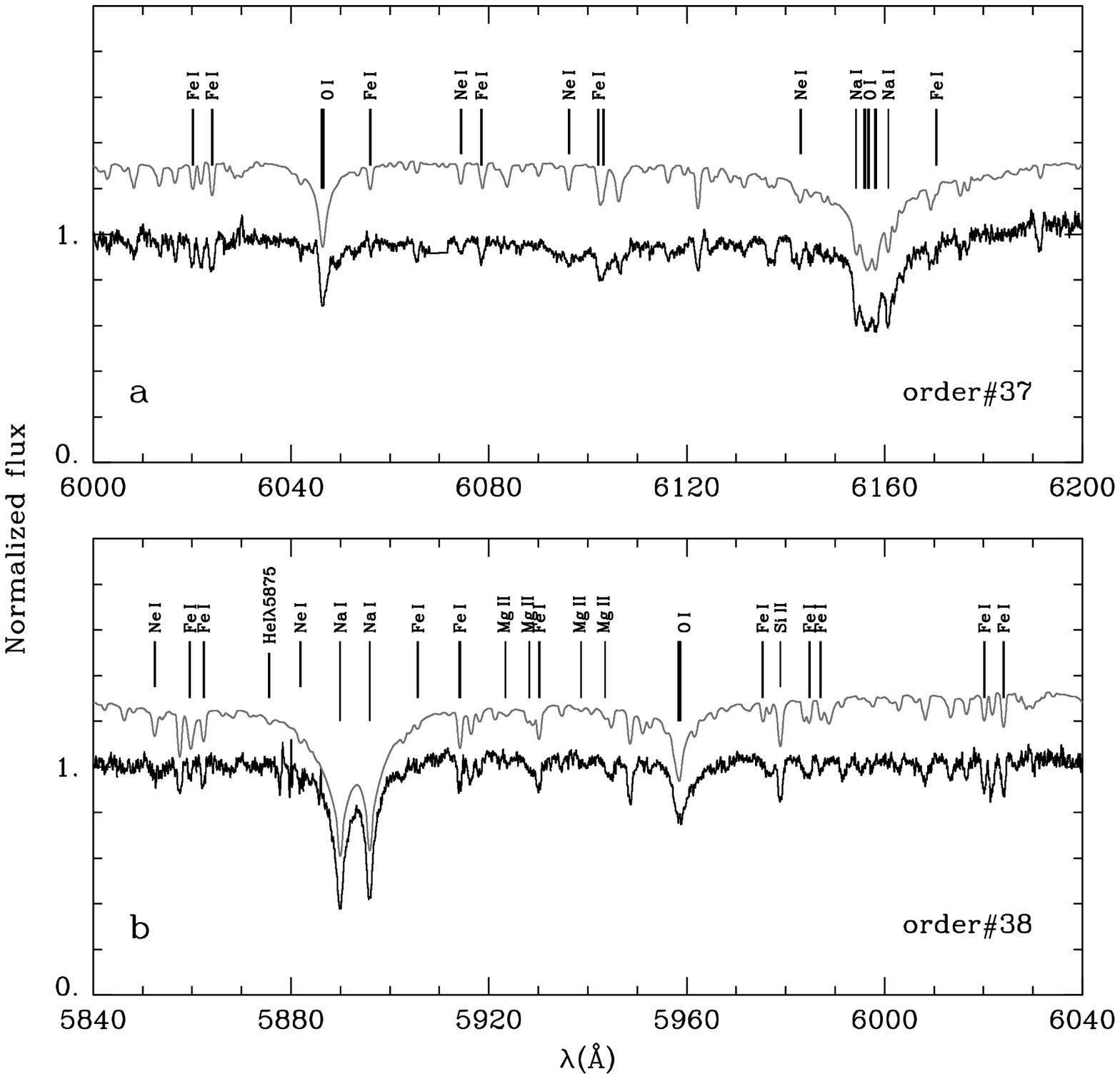}
\caption{{\bf High dispersion spectra and model.} Same as Fig.~\ref{fig_atl1} but for orders 37 and 38.
\label{fig_atl2}}
\end{figure}

\begin{figure}
\hspace{0cm}
\includegraphics[width=1.00\columnwidth]{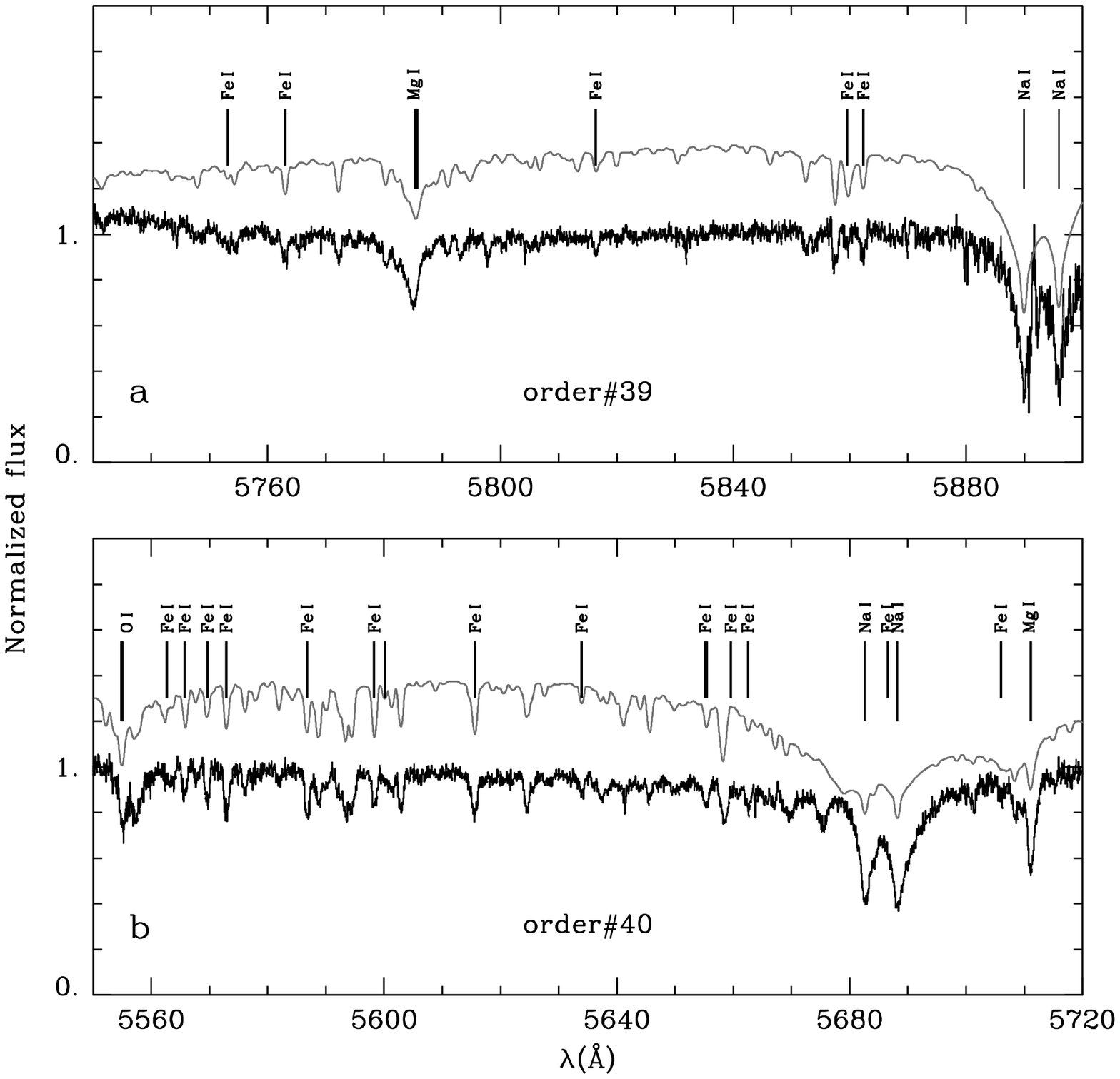}  
\caption{{\bf High dispersion spectra and model.} Same as Fig.~\ref{fig_atl1} but for orders 39 and 40.
\label{fig_atl3}}
\end{figure}

\begin{figure}
\hspace{0cm}
\includegraphics[width=1.00\columnwidth]{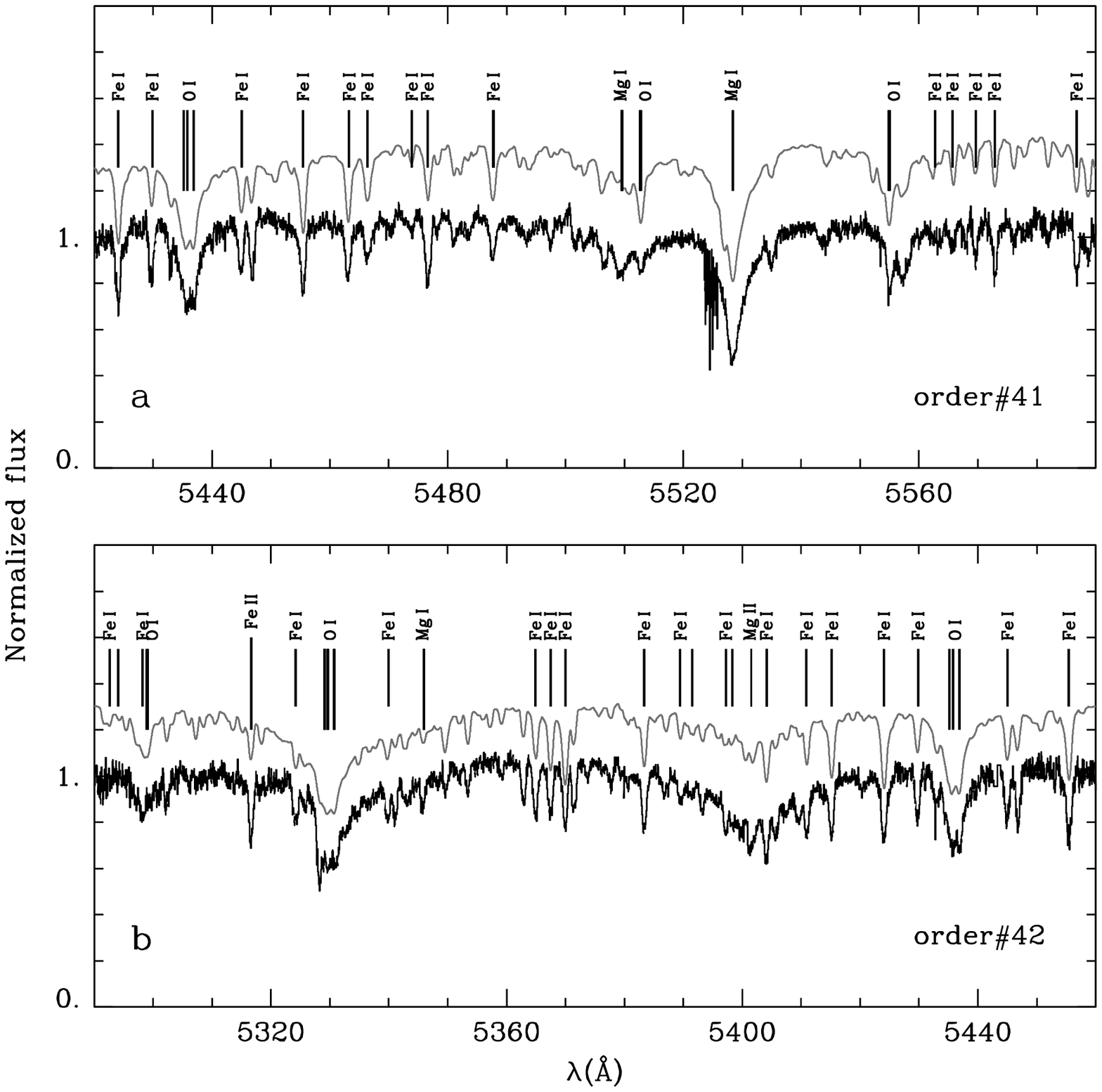}  
\caption{{\bf High dispersion spectra and model.} Same as Fig.~\ref{fig_atl1} but for orders 41 and 42.
\label{fig_atl4}}
\end{figure}

\begin{figure}
\hspace{0cm}
\includegraphics[width=1.00\columnwidth]{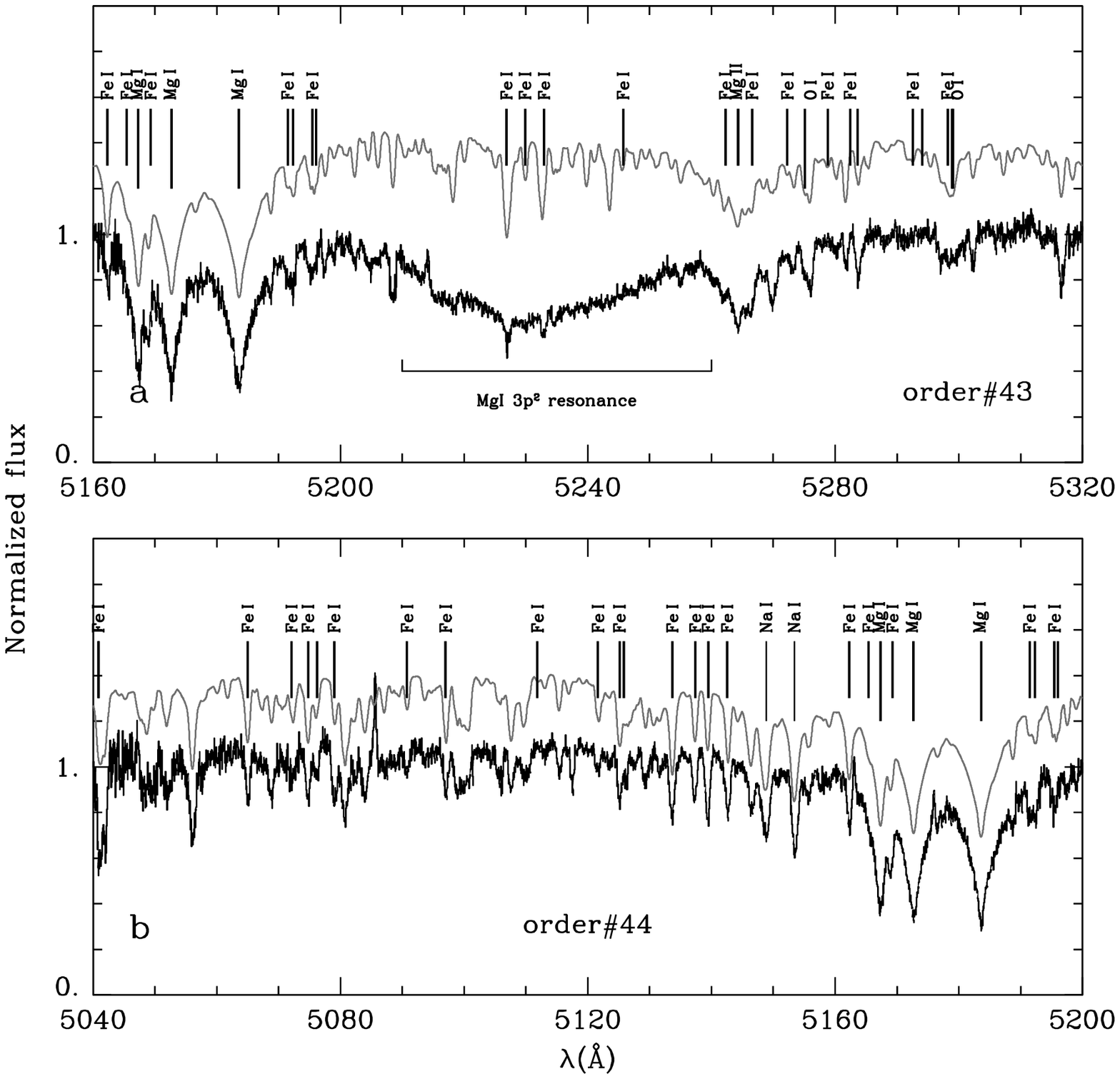}  
\caption{{\bf High dispersion spectra and model.} Same as Fig.~\ref{fig_atl1} but for orders 43 and 44.
\label{fig_atl5}}
\end{figure}

\begin{figure}
\hspace{0cm}
\includegraphics[width=1.00\columnwidth]{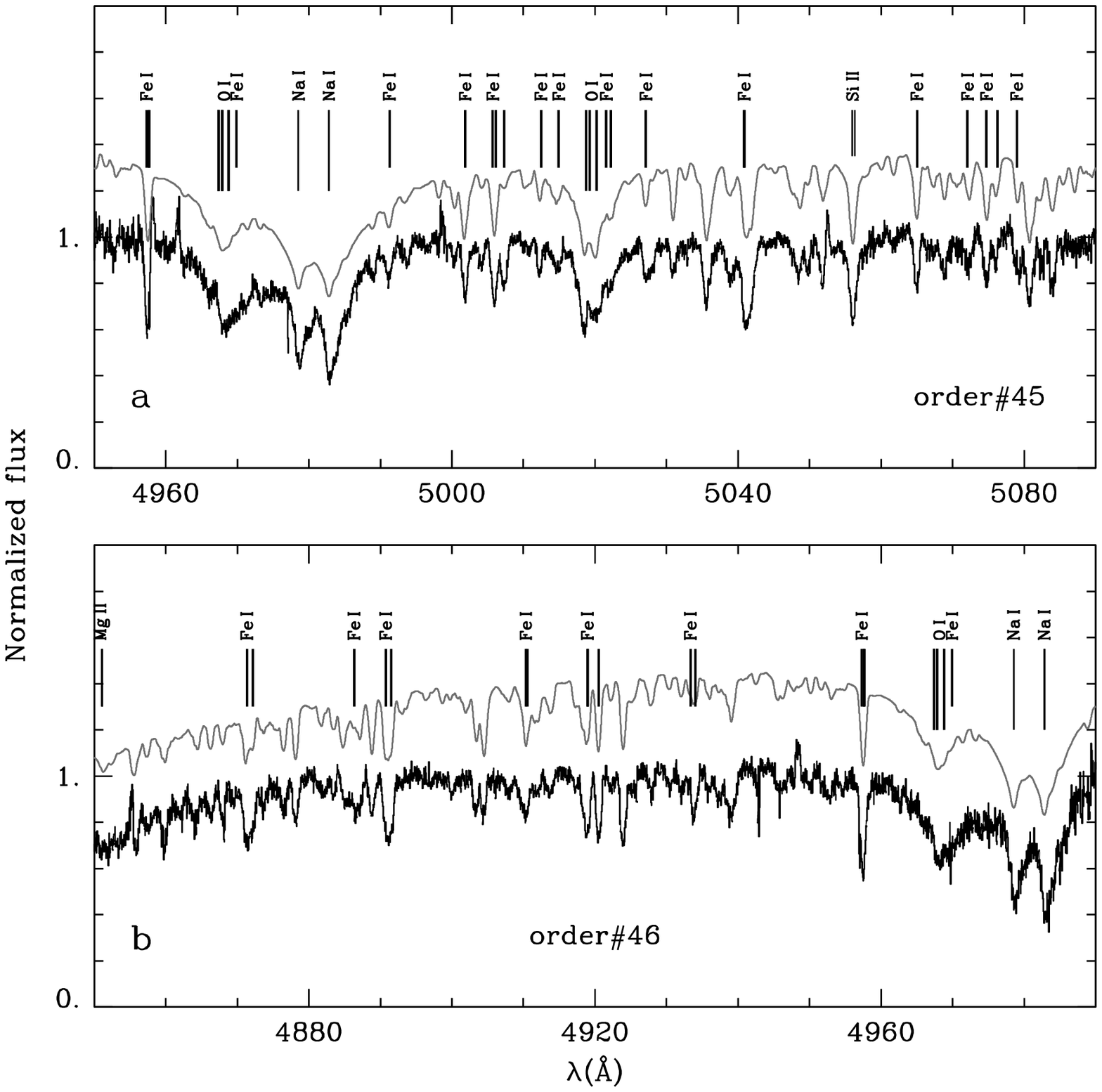}  
\caption{{\bf High dispersion spectra and model.} Same as Fig.~\ref{fig_atl1} but for orders 45 and 46.
\label{fig_atl6}}
\end{figure}

Figs.~\ref{fig_atl1}, \ref{fig_atl2}, \ref{fig_atl3}, \ref{fig_atl4},
\ref{fig_atl5}, and \ref{fig_atl6} show details of high-dispersion
spectra obtained with GRACES. The echelle spectral orders numbered from 34 to 46 cover
a range of 1900\AN\ between $\lambda=$4850 to 6750\AN\ and are compared to the
best-fitting spectral synthesis.
A broad feature near 5230\AN\ (Fig.~\ref{fig_spec}) is tentatively identified with a resonance in the Mg~{I}\,3p$^2$ photoionization cross-section. 
The available theoretical cross-section included in our spectral synthesis has
insufficient wavelength coverage and does not model the observed feature satisfactorily.
The spectral synthesis includes H$\alpha$ and He~{I}$\lambda5875$ with hydrogen and helium at
their respective abundance upper limit. 
Numerous lines of Fe~{I} are identified over the entire spectrum, particularly in the blue
($\approx 5000$\AN).
A few weak spectral lines among many detected in the GRACES spectra remain unidentified or are poorly reproduced by the
spectral line synthesis due to inaccurate atomic
data or unidentified line blends.\\

\noindent {\bf S3: Model atmosphere calculations, fitting techniques, and stellar parameters}\\

A grid based spectral modelling using a pre-calculated spectral library 
proved very inefficient due to the 
peculiar composition of LP~40-365 which would require a 
prohibitively large number of models to be computed.
Therefore we conducted our analysis 
with the iterative steepest-descent spectral
analysis package {\sc XTgrid} \cite{nem2012} featuring {\sc Tlusty} 
(version 204) 
\cite{hub1995} and {\sc Synspec} (version 49) \cite{lan2007}, to calculate 
tailor made model atmospheres and their corresponding synthetic spectra.
The atmosphere structure models and spectral synthesis have been performed
consistently including the first 30 elements of the periodic table. 
Helium and the astrophysically important light metals, alpha- and iron-group 
elements (C, N, O, Ne, Na, Mg, Al, Si, S, Ca, Fe, Ni) were 
considered in full-NLTE with detailed model atoms, while the rest of the
first 30 elements were included assuming LTE conditions. 
All elements contribute to the electron density, and hence the pressure.
Due to the low temperature and high metallicity the stellar plasma is dominated 
by electrons with contributions from weakly ionized metals, 
therefore collisions occur predominantly with electrons 
producing the Stark broadening.
Model atoms for low excited ions were taken from the {\sc Tlusty} web
page\cite{tlusty1}
and complemented with additional neutral and 
singly-ionized model atoms\cite{pri2003} and detailed
Mg~{III} and Na~{II} model atoms\cite{nem2010}.
We investigated the role of convective energy transport 
with the {\sc Tlusty} models and
found that convection is inefficient throughout the atmosphere. 

The {\sc XTgrid} procedure was initiated with a metal-rich generic 
extremely low mass WD model, and, by successive
approximations along the steepest gradient of the global $\chi^2$ surface,
it converged on a solution. 
Each model parameter, such as $T_{\rm eff}$, $\log{g}$, individual 
abundances, and the projected rotational velocity 
has been treated separately. 
To avoid trapping in local minima the procedure regularly increased 
the step sizes and re-started its descent.
The fitting procedure automatically adjusts the atomic data input to the 
actual model. 
By evaluating the ionization fractions it assigns the most detailed 
model atoms to the most populated ionization stages. 
The spectral analysis was based on the relative line strengths and line
profiles. 
We applied a piecewise normalization of the synthetic spectra together 
with a linear correction of the flux 
to match the theoretical continuum to the observed one. 
The lengths of these fitting ranges were carefully selected to be short around 
narrow lines and long enough to include the full profiles of strong 
lines. 
This approach also reduces the systematics caused by interstellar reddening. 
We found that the neutral and singly ionized ions, 
in particular Mg~{I}, Mg~{II}, Fe~{I}, Fe~{II}, Ca~{I},
and Ca~{II} are the most sensitive temperature indicators, 
while the surface gravity is constrained by the Stark broadening of the 
strongest lines.
For the spectral synthesis we use the line list available on the 
{\sc Synspec} web page and complemented it with the latest atomic 
transitions data 
compiled by Robert L. 
Kurucz\cite{kurucz1},
and complemented with oscillator strengths of 20\% accuracy or better
from the Atomic Spectra Database at the National Institute of Standards and
Technology (NIST)\cite{nist1}.

The uncertainties were estimated from the $\chi^2$
surface statistics. 
Models were calculated in one dimension until 
the reduced $\chi^2$ values reached the confidence limit for the given 
number of free parameters.  
Parameter correlations are not considered in the error analysis; the 
procedure exploits these during the iterative fitting, such that 
parameter correlations in the proximity of the solution are low. 
The number of independent variables ($p=15$) in the global fitting procedure 
accounts for the main stellar
parameters ($T_{\rm eff}$, $\log{g}$), abundant or spectroscopically dominant elements 
(He, C, N, O, Ne, Na, Mg, Si, Al, Ca, Fe, Ni), and the projected rotation velocity ($v_{\rm rot}\sin{i}$).
Error bars are estimated with the condition $\chi^2_{\alpha}=\chi^2_{\rm min}+\chi^2_p(\alpha)$,
where $\alpha$ is the significance. Adopting 90\% confidence level, i.e., $\alpha=0.1$ and
$\chi^2_{15}(0.1)=22.3$, we calculated such errors for the $T_{\rm eff}$, $\log{g}$, and $v_{\rm rot}\sin{i}$. 
We redetermined the individual abundance errors with 90\% confidence using the best-fitting 
model and keeping all other parameters fixed ($p=1$, $\chi^2_{1}(0.1)=2.7$) and adding
in quadrature the measured variations for each abundance during the global fitting procedure. 

\begin{figure}
\hspace{0cm}
\includegraphics[width=1.00\columnwidth]{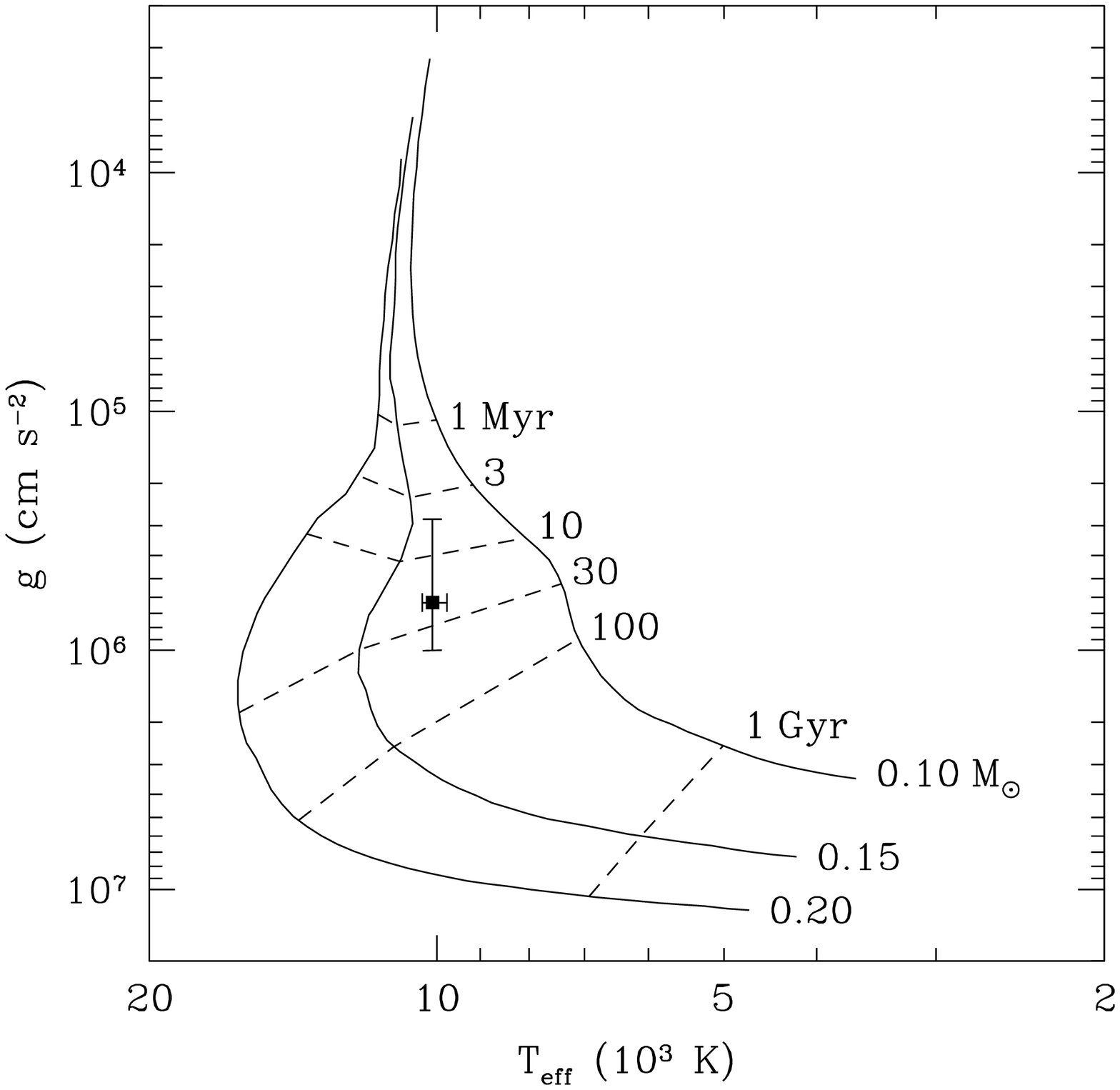}  
\caption{{\bf Evolutionary tracks for low mass white dwarfs.} The theoretical parameters, surface gravity ($\log{g}$) versus effective temperature ($T_{\rm eff}$), for stellar remnants are based on mass-radius relations \cite{alt1997} labelled with the mass (full lines) and cooling ages (dashed lines). The parameters of LP~40-365
are shown with a solid square including error bars.
\label{fig_tg}}
\end{figure}

Our analysis describes the surface properties of LP~40-365, but we have few constraints on
its internal structure. We rule out hydrogen and helium because any amount of these elements would quickly diffuse up and
contaminate the photospheric layers. The dominant photospheric components, oxygen and neon, are likely core material. Chemical
separation would also leave lighter elements such as oxygen and neon on top of heavier elements such as silicon and iron. 
Fig.~\ref{fig_tg} shows calculated stellar parameters based on mass-radius relations adopted to estimate the cooling age and mass of
the remnant\cite{alt1997}. Theoretical radii calculated assuming interior compositions
dominated by light elements (He, O, Si) but without
hydrogen envelopes vary by only a few percent above 0.3\msun\ \cite{pan2000} and the predicted radius
inflation is driven by finite temperature effects. A similar behavior is expected at lower
mass although it requires extrapolation. Models with an iron core and similarly low-mass
are not available but are expected to have, for a given mass, a smaller radius than their light-element counterpart.
However, the iron core fraction is unknown. Surface diffusion time scales are comparable to the cooling age ($\approx 10^7$ yr) of
LP~40-365\cite{paq1986} and without evolutionary computations including the effect of diffusion
it is not possible to determine, for example\cite{jor2012}, 
the fraction of the 0.02\msun\
of iron-group material that could have diffused at the center of the 0.28\msun\ CO-dominated core.

By interpolating the cooling tracks\cite{alt1997} depicted in Fig.~\ref{fig_tg} we converted the surface 
gravity and temperature measurements into an age of
$\approx 5-50$\,Myr and a mass of 0.13-0.15\msun. The radius and corresponding error bars were
estimated following $R=\sqrt{GM/g}$ and the radius error was estimated by
allowing $g$ and $M$ to vary within their quoted errors. Error bars for the absolute magnitude were
similarly propagated.\\

\noindent {\bf S4: Spectral energy distribution}\\

We obtained ultraviolet photometric measurements ($F_{\rm UV}$ and $N_{\rm UV}$) from the
{\it Galaxy Evolution Explorer} (\emph{GALEX}) all-sky survey\cite{mor2007}.
We also obtained IR photometric measurements from the Two Micron All Sky Survey (2MASS)
\cite{skr2006} and {\it Wide-field Infrared Survey Explorer} (\emph{WISE})
\cite{wri2010}.
We also include an $R$-band measurement from the Palomar Transient Factory (PTF) photometric catalog 1.0
converted to the Johnson-$R$ photometric band using $R\approx R_{\rm PTF}-0.14$\,mag\cite{ofe2012b}. The object was not
found to vary with $\sigma(R_{\rm PTF})<0.04$ mag\cite{ofe2012a}.
We also include the Sloan $g$ and $r$ and Johnson $V$ and $B$ measurements from the AAVSO
Photometric All Sky Survey (APASS) data release number 9 (DR9)\cite{wat2016}.
Table~\ref{tbl_phot} lists the measurements calibrated with either the
Vega or AB system as prescribed.

\begin{figure}
\hspace{0cm}
\includegraphics[width=1.00\columnwidth]{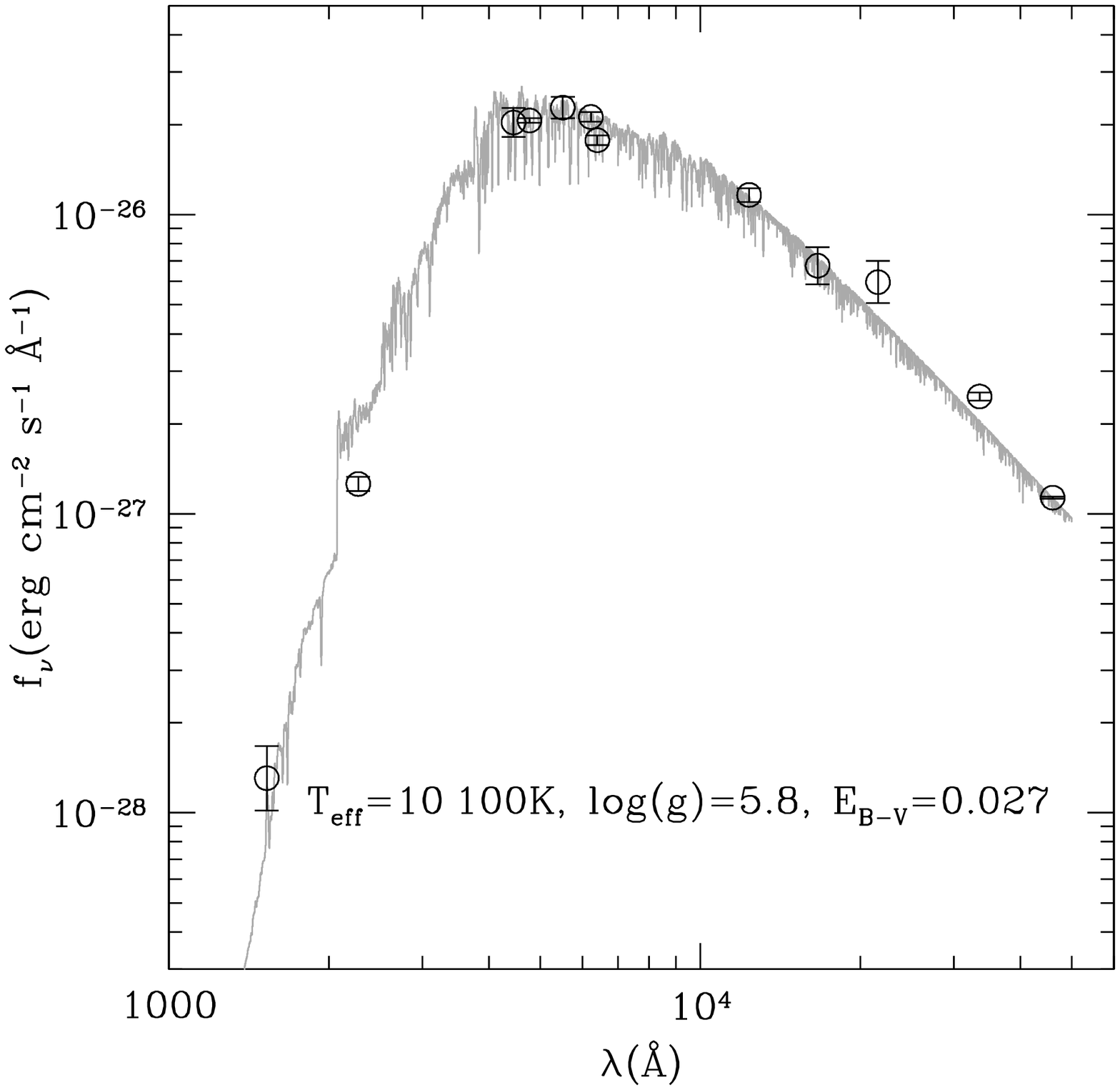}
\caption{{\bf Spectral energy distribution of LP~40-365.} The ultraviolet, optical, and infrared photometric measurements (Table~\ref{tbl_phot}) are shown with open circles and expressed as the
flux measured at Earth ($f_\nu$) versus wavelength ($\lambda$) and are compared to a 
model spectrum (grey line).
The adopted stellar parameters ($T_{\rm eff}$, $\log{g}$) and the reddening coefficient ($E_{B-V}$) 
applied to the model
spectrum are shown.
\label{fig_sed}}
\end{figure}

The peak of the spectral energy distribution (Fig.~\ref{fig_sed}) is well reproduced by a spectral synthesis based on parameters
measured with the optical spectral line
analysis. The concordance validates our conclusions on the properties of this object.
Since LP~40-365 lies well outside the Galactic plane, we computed the model spectrum including the effect of
the total extinction measured in the line-of-sight, $E_{B-V}=0.027$\cite{sch1998}.
The synthetic optical magnitudes $(B,V,R,g,r)=(15.73,15.57,15.51,15.60,15.62)$\,mag computed using the model spectrum are in agreement with the measurements
(Table~\ref{tbl_phot}) with the exception of $R_{\rm PTF}$ which is marginally consistent.
The synthetic ultraviolet magnitudes $(F_{\rm UV},N_{\rm UV}) = (21.34,18.01)$ differ from the observed magnitudes although the $F_{\rm UV}$ measurement
is uncertain. The discrepancy in the $N_{\rm UV}$ magnitudes ($-0.6$ mag) is possibly caused by
missing ultraviolet opacities in the model calculations that would affect the spectral energy distribution.\\

\noindent {\bf S5: Stellar kinematics}\\

For the computation of the orbit of LP~40-365 we have used a code
based on the Galactic potential of Allen \& Santillan\cite{all1991}. In a cylindrical
coordinate system with origin in the Galactic center this potential is
analytical, time-independent and symmetric with respect to the
z-axis. This potential consists of three components: a disk and a
central bulge of the Miyamoto-Nagai\cite{miy1975} form and a massive spherical halo
extending to a radius of 100\,kpc. The distance of the Sun from the
Galactic center is taken to be 8.5\,kpc. The local
circular velocity for this Galactic potential is $V_0=220.0$\kmps\ and the local escape
velocity is $V_{\rm esc}=535.7$\kmps.

We have integrated the equations of motion using a fourth order
Runge-Kutta algorithm with integration time steps chosen to ensure
that the total energy $E_{\rm tot}$ is conserved with a precision of
at least $\Delta E_{\rm tot}/E<10^{-10}$. The possible orbits of
LP~40-365 were then integrated backwards in time over 120\,Myr, which
largely exceeds the longest possible age that we have estimated for LP~40-365.

\section*{Acknowledgments}

Based on observations obtained at
the Gemini Observatory, programme GN-2016A-FT-24, using ESPaDOnS, located at the
Canada-France-Hawaii Telescope (CFHT). CFHT is operated by the National Research
Council of Canada, the Institut National des Sciences de l'Univers of the Centre
National de la Recherche Scientique of France, and the University of Hawaii. The
operations at the Canada-France-Hawaii Telescope are conducted with care and respect
from the summit of Maunakea which is a significant cultural and historic site.
ESPaDOnS is a collaborative project funded by France (CNRS, MENESR, OMP, LATT),
Canada (NSERC), CFHT and ESA. ESPaDOnS was remotely controlled from the Gemini
Observatory, which is operated by the Association of Universities for Research in
Astronomy, Inc., under a cooperative agreement with the NSF on behalf of the Gemini
partnership: the National Science Foundation (United States), the National Research
Council (Canada), CONICYT (Chile), Ministerio de Ciencia, Tecnolog\'{i}a e Innovaci\'{o}n
Productiva (Argentina) and Minist\'{e}rio da Ci\^{e}ncia, Tecnologia e Inova\c{c}\~{a}o (Brazil).
Based on observations obtained at Kitt Peak National Observatory, National Optical
Astronomy Observatory, 
which are operated by the Association of Universities for
Research in Astronomy under cooperative agreement with the National Science Foundation.

This publication makes use of data products from the Wide-field Infrared Survey
Explorer, 
which is a joint project of the University of California, Los
Angeles, and the Jet Propulsion Laboratory/California Institute of Technology,
funded by the National Aeronautics and Space Administration.
This publication makes use of data products from the Two Micron All Sky Survey,
which is a joint project of the University of Massachusetts and the Infrared
Processing and Analysis Center/California Institute of Technology, funded by
the National Aeronautics and Space Administration and the National Science
Foundation.
This publication makes use of data products from the AAVSO
Photometric All Sky Survey (APASS). 
Funded by the Robert Martin Ayers
Sciences Fund and the National Science Foundation.

The Gemini data are available
at\ \burl{https://archive.gemini.edu/searchform/NotFail/notengineering/cols=CTOWEQ/GN-2016A-FT-24}.
The KPNO and WHT data may be retrieved at\ \burl{http://archive.noao.edu/search/query}\ (program 2015A-0102)
and\ \burl{http://casu.ast.cam.ac.uk/casuadc/ingarch/query}\ (program OPTICON 2015A/22), respectively, and specifying the object coordinates.
All photometric data may be retrieved at\ \burl{http://vizier.u-strasbg.fr/viz-bin/VizieR} using the object coordinates.
The MDM data are available at\ \burl{https://zenodo.org/record/826288} (DOI: 10.5281/zenodo.826288).
{\sc Tlusty} and {\sc Synspec} are public codes available at\ \burl{http://nova.astro.umd.edu/} and the
best-fitting spectral synthesis for LP40-365 is available at\ \burl{https://zenodo.org/record/826139} (DOI: 10.5281/zenodo.826139).
{\sc XTgrid} is available through\ \burl{https://astroserver.org/}.

\clearpage

\clearpage 

\begin{table}
\centering
\caption{{\bf Barycentric radial velocity measurements.} The spectra
were obtained at four observatories at the heliocentric julian date (HJD)
and were used to measure the radial velocity ($v_{\rm r}$) and the uncertainty
on the velocity ($\sigma_{v_{\rm r}}$). Barycentric velocities are
corrected to the barycenter of the solar system.
\label{tbl_vel}}
\renewcommand{\arraystretch}{0.8}
\begin{tabular}{lccr}
\\
\hline
Observatory & HJD & $v_{\rm r}$ & $\sigma_{v_{\rm r}}$ \\
     & 2450000+ & (\kmps) & (\kmps) \\
\hline
KPNO & 7075.0110   & 515.8 & 11.8 \\
     & 7075.0340   & 484.9 & 11.5 \\
WHT  & 7189.4283   & 499.0 & 2.0 \\
     & 7189.6349   & 505.5 & 4.8 \\
     & 7189.6757   & 505.1 & 3.9 \\
     & 7190.5045   & 498.8 & 3.4 \\
     & 7190.6606   & 502.1 & 3.0 \\
MDM  & 7197.6506   & 497.4 & 18.8 \\
     & 7197.6593   & 489.1 & 19.4 \\
     & 7198.6560   & 470.5 & 27.1 \\
     & 7198.6668   & 463.0 & 15.3 \\
     & 7205.6543   & 514.0 & 13.8 \\
     & 7205.6652   & 515.0 & 24.5 \\
     & 7404.0345   & 486.4 & 11.3 \\
     & 7404.0453   & 489.0 & 14.3 \\
     & 7404.0561   & 485.7 & 17.3 \\
     & 7404.0686   & 493.3 & 17.6 \\
     & 7462.8685   & 488.4 & 12.0 \\
     & 7462.8772   & 499.1 & 10.7 \\
Gemini-N & 7540.7796   & 493.1 &  2.8 \\
       & 7543.7854   & 492.6 &  2.8 \\
\hline
\end{tabular}
\end{table}

\begin{table}
\centering
\caption{{\bf Photometric measurements.} These measurements are used to plot the spectral
energy distribution shown in Fig.~\ref{fig_sed}. The conversion into flux units is performed
according to the relevant photometric calibration system (AB or Vega).}
\label{tbl_phot}
\begin{tabular}{lcc}
\\
\hline
Band & Measurement & System \\
\hline
$F_{\rm UV}$       & 21.11$\pm$0.27 mag & AB \\
$N_{\rm UV}$       & 18.65$\pm$0.06 mag & AB \\
$B$                & 15.80$\pm$0.12 mag & Vega \\
$V$                & 15.51$\pm$0.09 mag & Vega \\
$g$                & 15.61$\pm$0.02 mag & AB \\
$r$                & 15.58$\pm$0.04 mag & AB \\
$R$                & 15.60$\pm$0.04 mag & Vega \\
$J$                & 15.34$\pm$0.06 mag & Vega \\
$H$                & 15.47$\pm$0.15 mag & Vega \\
$K$                & 15.12$\pm$0.18 mag & Vega \\
$W1$               & 15.23$\pm$0.03 mag & Vega \\
$W2$               & 15.41$\pm$0.07 mag & Vega \\
\hline
\end{tabular}\\
\end{table}

\end{document}